\documentclass[aps,twocolumn,showpacs,amssymb]{revtex4}
\usepackage[all]{xy}
\usepackage{graphicx}
\usepackage{epsfig}
\begin{document}
%%%%%%%%%%%%%%%%%%%%%%%%%%%%%%%%%%%%%%%%%%%%%%%%%%%%%%%%%%%%%%%

\title{Neutrino Decay and  Neutrinoless Double Beta Decay in a 3-3-1 Model }
%%%%%%%%%%%%%%%%%%%%%%%%%%%%%%%%%%%%%%%%%%%%%%%%%%%%%%%%%%%%%%%%

\author{Alex G. Dias,$^{a}$ A. Doff,$^b$  C. A. de S. Pires,$^c$ P. S. Rodrigues da Silva$^{c}$}
\affiliation{$^a${\small Instituto de  F\'{\i}sica, Universidade
de S\~ao Paulo,\\ Caixa Postal 66.318, 05315-970,S\~ao Paulo-SP, Brazil}\\
$^b${\small Instituto de  F\'{\i}sica Te\'{o}rica, Universidade
Estadual Paulista, Rua Pamplona 145,
01405-900 S\~{a}o Paulo - SP, Brazil}\\
$^c${\small  Departamento de F\'{\i}sica, Universidade Federal da
Para\'{\i}ba, Caixa Postal 5008, 58051-970, Jo\~ao Pessoa - PB,
Brazil.}}

\date{\today}
\begin{abstract}
In this work we show that  the implementation of spontaneous breaking of the lepton number in the 3-3-1 model with right-handed neutrinos gives rise to fast neutrino decay with majoron emission and generates a bunch of new contributions to the  neutrinoless double beta decay.
\end{abstract}

\maketitle

%%%%%%%%%%%%%%%%%%%%%%%%%%%%%
\section{Introduction}
The  detection of neutrino oscillation\cite{SK} experimentally indicates that neutrinos are massive particles and that flavor lepton number is not  conserved.  Since the standard model has no room neither for massive neutrinos, and consequently, nor for flavor lepton number violation, thus the neutrino oscillation  experiments provided an important and undoubted signal of new physics.

Despite of the present experimental advance in neutrino physics, we do not know yet if neutrinos are Dirac or Majorana particles. The latter case is particularly interesting because demands  violation of the lepton number which has important consequences in particle physics. 

The natural way of getting violation of the lepton number is through its spontaneous breaking. However, when the lepton number is not a gauge symmetry, the Goldstone boson that results in spontaneous breaking of the lepton number (SBLN), the majoron, is very problematic, as this scalar has also to be invisible, i.e., very weakly interacting with ordinary matter.

It was recently shown in Ref.~\cite{majoron} that  the $SU(3)_C\otimes SU(3)_L\otimes U(1)_N$ (3-3-1 for short) model with right-handed neutrinos\cite{model} disposes automatically  of a  structure that makes the SBLN possible once the associated Goldstone boson is an invisible majoron\cite{model2}. 

The 3-3-1 model with right-handed neutrinos is one of the possible models allowed by the 3-3-1 gauge symmetry where  the fermions are distributed in the following representation content\cite{model}. Leptons come in triplets and singlets
\begin{eqnarray}
f_{aL} = \left (
\begin{array}{c}
\nu_a \\
e_a \\
\nu^{c}_a
\end{array}
\right )_L\sim(1\,,\,3\,,\,-1/3)\,,\,\,\,e_{aR}\,\sim(1,1,-1),
 \end{eqnarray}
where $a = 1,\,2,\, 3$ refers to the three families. In the quark sector, one generation comes in the triplet and the other two compose
an anti-triplet with the following content,
\begin{eqnarray}
&&Q_{iL} = \left (
\begin{array}{c}
d_{i} \\
-u_{i} \\
d^{\prime}_{i}
\end{array}
\right )_L\sim(3\,,\,\bar{3}\,,\,0)\,,u_{iR}\,\sim(3,1,2/3),\,\,\,\nonumber \\
&&\,\,d_{iR}\,\sim(3,1,-1/3)\,,\,\,\,\, d^{\prime}_{iR}\,\sim(3,1,-1/3),\nonumber \\
&&Q_{3L} = \left (
\begin{array}{c}
u_{3} \\
d_{3} \\
u^{\prime}_{3}
\end{array}
\right )_L\sim(3\,,\,3\,,\,1/3),u_{3R}\,\sim(3,1,2/3),\nonumber \\
&&\,\,d_{3R}\,\sim(3,1,-1/3)\,,\,u^{\prime}_{3R}\,\sim(3,1,2/3)
\label{quarks} 
\end{eqnarray}
where $i=1,2$. The primed quarks
are the exotic ones but with the usual electric charges.

In the scalar sector, this model possess three triplets responsible for the fermion masses,
\begin{eqnarray}
\eta = \left (
\begin{array}{c}
\eta^0 \\
\eta^- \\
\eta^{\prime 0}
\end{array}
\right ),\,\rho = \left (
\begin{array}{c}
\rho^+ \\
\rho^0 \\
\rho^{\prime +}
\end{array}
\right ) ,\, \chi = \left (
\begin{array}{c}
\chi^0 \\
\chi^{-} \\
\chi^{\prime 0}
\end{array}
\right ) , \label{scalarcont} 
\end{eqnarray}
both transforming as  $\eta \sim ({\bf 1}\,,\,{\bf 3}\,,\,-1/3)$ and $\chi \sim ({\bf 1}\,,\,{\bf 3}\,,\,-1/3)$
and $\rho \sim ({\bf 1}\,,\,{\bf 3}\,,\,2/3)$.

In the gauge sector, the model recovers the standard gauge bosons  and disposes of five more called  $V^{\pm}$, $U^0$, $U^{0 \dagger}$ and $Z^2$~\cite{model,phenomenology}. The interactions of the  first four gauge bosons with the leptons are
\begin{eqnarray}
	\frac{g}{\sqrt{2}}\bar e_L \gamma^\mu \nu^C_L V^+_\mu + \frac{g}{\sqrt{2}}\bar \nu_L \gamma^\mu \nu^C_L U^0_\mu + \mbox{H.c}.
	\label{g-b-interactions}
\end{eqnarray}
This means that the new gauge bosons $V^{+}$ and  $U^0$ both carry two units of lepton number, thus we call them vector bileptons.

On the other hand, these new gauge bosons present the following interactions with quarks
\begin{eqnarray}
&&\frac{g}{\sqrt{2}}\left( \bar u^{\prime}_{3L} \gamma^\mu d_{3L} + \bar u_{iL}\gamma^\mu d^{\prime}_{iL} \right) V^+_\mu \nonumber \\
&&\frac{g}{\sqrt{2}}\left(\bar u_{3L} \gamma^\mu u^{\prime}_{3L} + \bar d^{\prime}_{iL}\gamma^\mu d^{\prime}_{iL}\right)U^0_\mu +\mbox{H.c}.
\label{q-g-b-interac}
\end{eqnarray}
Consequently the new quarks, $u^{\prime}_3$  and $d^{\prime}_i$, also carry two units of lepton number, thus we call them  leptoquarks.

In order to see which other particles carry two units of lepton number, we have to look at the Yukawa interactions of the fermions with the scalar content of the model
\begin{eqnarray}
	{\cal L}^Y &=&\lambda^1_{ij}\bar Q_{iL}\chi^* d^{\prime}_{jR}+\lambda^2_{33}\bar Q_{3L}\chi u^{\prime}_{3R}+ \lambda^3_{ia}\bar Q_{iL}\eta^* d_{aR}+\nonumber \\
&&\lambda^4_{3a}\bar Q_{3L}\eta u_{aR}+ \lambda^5_{ia}\bar Q_{iL}\rho^* u_{aR}+\lambda^6_{3a}\bar Q_{3L}\rho d_{aR}+\nonumber \\
&& h_{ab}\bar f_{aL} \rho e_{bR}+\mbox{H.c}.
\label{yukawa}
\end{eqnarray}
Perceive that these interactions impose that the scalars $\eta^{\prime
0}$, $\rho^{\prime +}$, $\chi^ 0$ and $\chi^-$ also carry two units of lepton number, they are the scalar bileptons. 

In summary, we have the following lepton number assignment\cite{app} 
\begin{eqnarray}
&&
 {\mbox L}(V^+\,,\, U^{\dagger0}\,,\, u^{\prime}_{3} \,,\, \eta^{\prime
0}\,,\,\rho^{\prime +})=-2 ,\nonumber \\
&&{\mbox L}(V^- \,,\,U^0 \,,\,d^\prime_{i}
\,,\, \chi^ 0\,,\, \chi^-)=+2. \label{leptonnumber} \end{eqnarray}

Note that we have two neutral scalar carrying lepton number, $\eta^{\prime
0}$  and $\chi^ 0$. Therefore when one (or both) of these scalars develop a vacuum expectation value (VEV),
we are going to have spontaneous breaking of the lepton number. For sake of simplicity, in the mechanism of SBLN  developed in ref.~\cite{majoron} only $\eta^{\prime
0}$ developed VEV, $\eta^{\prime
0}\rightarrow \frac{\sqrt{2}}{2}\left( v_{\eta^{\prime}}+ R_{\eta^{\prime}}+iI_{\eta^{\prime}}\right)$. It was shown in Ref.~\cite{majoron} that  $I_{\eta^{\prime}}$ is the Majoron, $J$, and that it decouples from the other scalars. Based on the Yukawa interactions above, the Majoron does not couple directly with any lepton. Interesting enough, it  interacts only with quarks and leptoquarks,
\begin{eqnarray}
i\lambda^4_{3a}\bar u^{\prime}_{3L} u_{aR}J-i\lambda^3_{ia}\bar d^{\prime}_{iL} d_{aR}J.
\label{maj-quark-int}	
\end{eqnarray}

Despite of the fact that neutrinos do not present couplings with the majoron at tree level, such couplings will arise through one-loop corrections. In order to see that, consider the following interactions
\begin{eqnarray}
	h_{ab} \bar\nu^C_{aL}e_{bR}\rho^{\prime +} + h_{ab} \bar\nu^C_{bR}(e^C)_{aL} \rho^+  +i\lambda_9 v_\eta J \rho^+\rho^{\prime -},                                            
    \label{Yukawa-majorn-coupligs}
    \end{eqnarray}
 The first two terms come from the Yukawa interactions above and the last term comes from the potential given in Ref.~\cite{majoron}.

The three terms in Eq.(\ref{Yukawa-majorn-coupligs}) generate, through one-loop,  the following effective coupling among neutrinos and Majoron 
\begin{eqnarray}
	G_{ab}\bar\nu^C_{aR} \nu_{bL}J
	\label{effectiveJnu}
\end{eqnarray}
where 
\begin{eqnarray}
	G_{ab}\approx \frac{h_{ai} m_i h_{ib} \lambda_9 v_\eta}{16\pi^2m^2_{\rho^{\prime +}}},
	\label{couplingeffective}
\end{eqnarray}
with $a\,,\,b=1\,\,,2\,\,,3$  and $i=e\,\,,\mu\,\,,\tau$. $v_\rho$  and $v_\eta$ are the VEV´s of $\rho^0$  and $\eta^0$, respectively. 

Apart from the prediction of a viable Majoron, the mechanism of SBLN led to an interesting interplay among neutrino and charged lepton masses, as showed in Ref.~\cite{majoron}
\begin{eqnarray}
&& m^l_{ab}=h_{ab}v_{\rho},\nonumber \\
&& m^\nu_{ab} \approx \frac{ h_{ai} m_i h_{ib} \lambda_9v_\eta v_{\eta^{\prime}}}
    {16 \pi^2m^2_{\rho^{\prime+}}}.
    \label{leptonmass}
\end{eqnarray}
 This neutrino mass matrix originated from one-loop radiative corrections, but observe that  it can be obtained from the effective neutrino-majoron coupling by replacing $J$ by $v_{\eta^{\prime}}$.

 Thus we have a model that violates lepton number, disposes of  light Majorana neutrinos and contains a Majoron in its spectrum of scalars.  The investigation of neutrino decay  and neutrinoless double beta decay becomes mandatory in a model with such ingredients. This is the aim of this work. Basically we will study neutrino decay with Majoron emission and obtain the dominant contributions of the model for the neutrinoless double beta decay. What justify such investigation is the cosmological importance of fast neutrino decays and the role of the neutrinoless double beta decay in determining the nature of the neutrinos.

We organize our study in the following sequence. In Sec.~\ref{sec1} we derive the charged gauge boson mixing. Next in  Sec.~\ref{sec2} we obtain the lifetime for neutrino decaying with Majoron emission and in Sec.~\ref{sec3} we present  the significant contributions to the neutrinoless double beta decay. We finish our work with a brief conclusion in Sec.~\ref{sec4}.

\section{Charged gauge boson sector}
\label{sec1}

The main consequence of $v_{\eta^{\prime}}\neq 0$ in the charged gauge boson sector is a mixing among $W^+$  and the bilepton $V^+$. In order to  see that,  we must open the scalars kinetic term, $\displaystyle\sum_{\varphi}^{\eta,\rho,\chi} \left ( {\cal D}_\mu \varphi \right)^{\dagger} \left( {\cal D}^\mu \varphi \right)$, then obtaining,
\begin{eqnarray}
\left( W^+_\mu\,\,\,\,\,\,  V^+_\mu \right) \frac{g^2}{4}
\left (
\begin{array}{lcr}
v^2_\rho + v^2_\eta & v_\eta v_{\eta^{\prime}} \\
v_\eta v_{\eta^{\prime}} & v^2_\chi + v^2_\rho +v^2_{\eta^{\prime}} 
\end{array}
\right )\left (
\begin{array}{c}
W^{-\mu} \\
V^{-\mu} 
\end{array}
\right ).
\label{bileptonmix}
\end{eqnarray}
In view of this mixing, the new charged gauge boson mass eigenstates, that we call $\tilde{W}^+$  and $\tilde{V}^+$, relate to  $W^+$  and $V^+$ in the following way,
\begin{eqnarray}
\left (
\begin{array}{c}
\tilde{W}^+_{\mu} \\
\tilde{V}^+_{\mu} 
\end{array}
\right )=\left (
\begin{array}{lcr}
c_\theta & -s_\theta \\
s_\theta & c_\theta 
\end{array}
\right )\left (
\begin{array}{c}
W^+_{\mu} \\
V^+_{\mu} 
\end{array}
\right ),
\label{chargedmix}
\end{eqnarray}
where $\tan\theta\approx \frac{v_{\eta^{\prime}}v_\eta}{v^2_\chi}$. According with our notation, $c_\theta=\cos\theta$ and $s_\theta=\sin\theta$.
 
As a consequence of this mixing,  each $\tilde{W}^+$ and $\tilde{V}^+$  will interact, separately,  with the two leptonic charged currents of the model, $j^\mu=\frac{g}{\sqrt{2}}\bar e_L U\gamma^\mu \nu_L$  and $J^{C \mu}=\frac{g}{\sqrt{2}}\bar e_L U\gamma^\mu \nu^C_L$, as shown below,
\begin{eqnarray}
{\cal L}^{CC}_l =\left( j^\mu c_\theta +j^{C\mu}s_\theta \right) \tilde{W}^+_\mu +\left(j^{C\mu}c_\theta -j^\mu  s_\theta  \right) \tilde{V}^+_\mu ,
\label{newCC}
\end{eqnarray} 
where $U$ is the PMNS leptonic mixing matrix.

In the quark sector, the standard quark current now interact with  both $\tilde{W}^+$ and $\tilde{V}^+$ in the following way 
\begin{eqnarray}
{\cal L}^{CC}_q =\frac{g}{\sqrt{2}}\bar u_L V_{CKM}\gamma^\mu d_L	\left(  \tilde{W}^+_\mu c_\theta + \tilde{V}^+_\mu s_\theta \right),
\label{quarCC}
\end{eqnarray}
with $V_{CKM}$ is the usual Cabibo-Kobayashi-Maskawa mixing between quarks.

It is in this  new basis of charged current interactions  that we will investigate neutrino decay with Majoron emission and neutrinoless double beta decay.

\section{Neutrino decay with Majoron emission}
\label{sec2}
The discovery of neutrino oscillation\cite{SK} by atmospheric and solar neutrino experiments can be also used as a definitive proof that neutrinos are massive particles. The surprise of those discoveries were twofold. The lightness of the neutrino masses and the largeness of their mixing.  The present data provides (at 3$\sigma$ level)\cite{atm,solar,reactor,global} $\Delta m^2_{31}= (1.6-3.6) \times 10^{-3}$eV$^2$  and  angle $\theta_{23}\cong45$ degrees for atmospheric neutrino oscillation, while for solar neutrinos we have $\Delta m^2_{21}= (6.8 - 8.1) \times 10^{-5}$eV$^2$  and  angle $\theta_{12}\cong 30,9 - 34,2$ degrees. For the third angle we have an upper bound $\theta_{13}\leq 6$ degrees.  Despite these  tiny differences among neutrino masses, its decay is possible through appropriate modes. 
%%%%%%%%%%%%%%%%%%%%%%%%%%%%%%%%%%%%%%%%%%%%%%%%%%%%%%%%%%%%%%%%%%%%%%%%%%%%%%%%%
%              Figura decaimento de neutrino L.H, formato EPS. 
%%%%%%%%%%%%%%%%%%%%%%%%%%%%%%%%%%%%%%%%%%%%%%%%%%%%%%%%%%%%%%%%%%%%%%%%%%%%%%%%%
\begin{figure}[ht]
\begin{center}
\epsfig{file=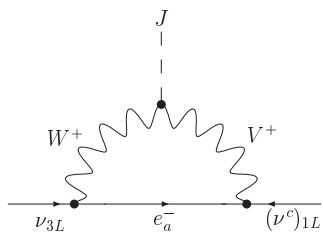,width=0.25\textwidth}
\caption{One loop diagram that leads to left-handed neutrino decay.}
\end{center}
\end{figure}

But before we start our analysis of neutrino decay with Majoron emission, remember that  the known Majorons  are singlet, doublet or triplet Majoron\cite{singlet,doublet,triplet}  under the $SU(3)_C\times SU(2)_L\times U(1)_Y$ symmetry. In the singlet case, the Majoron couples directly only to heavy right-handed neutrinos. In this case the width of the decay is suppressed by the scale of the heavy right-handed neutrino mass, which usually is at the GUT scale. Consequently this leads to a large neutrino lifetime\cite{moha}.  On the other hand, Majorons that belong to doublets or triplets do couple directly with left-handed neutrinos and then can lead to fast neutrino decay~\cite{moha}. However these Majorons were already ruled out by experiments\cite{concha}. 

Let us return to our model. There are two basic differences among our Majoron and the other known Majorons.  First,  our Majoron belongs to a triplet, but now under the 3-3-1 symmetry. Second, instead of interacting with leptons, it interacts with leptoquarks. This implies that the neutrino decay process, $\nu_i \rightarrow \nu_j +J$,  is only possible through radiative corrections. Besides,  this process happens through two modes. In one mode the daughter neutrino is a left-handed one, while in the other mode it is a right-handed neutrino. In view of all these differences, it turns out necessary to establish if   neutrinos are stable or not in our model, which can have important consequences for cosmology.

We start studying the mode $\nu_ {iL} \rightarrow (\nu^C)_{JL}+J$. This mode is  allowed by the charged currents in Eq. (\ref{newCC}) and the following interaction among the Majoron and the charged gauge bosons originated from the kinetic term  $({\cal D}_\mu \eta)^{\dagger}({\cal D}^\mu \eta)$,
\begin{eqnarray}
	i\frac{g^2}{2}v_\eta W^{+\mu} V^-_\mu J.
\label{W+V+J}
\end{eqnarray}
We restrict our analysis to the main contributions to the  neutrino decay processes. In view of this, in this section  we neglect the mixing in Eq. (\ref{chargedmix}). Moreover, since we are interested in fast neutrino decay, we also restrict our study  to the decay $\nu_{3 L} \rightarrow (\nu^C)_{1 L}+J$ where $\nu_3$  and $\nu_1$ are the neutrino mass eigenstates. This process is depicted in FIG.~(1). 

After summing over all the intermediate charged leptons ,and assuming  $m_{\nu_3} > m_{\nu_1}$, we obtain for the decay width in the rest frame,
\begin{eqnarray}
	\Gamma_{\nu_{3L} \rightarrow (\nu^C)_{1L}J}&&=\frac{ g^8(\displaystyle\sum_{a}U_{3 a}U_{a 1})^2 v_\eta^2 m^3_{\nu_3} }{2 (64)^3\pi^5(m^2_{V}-m^2_{W})^2}\nonumber \\
	&&\times\left( 1-\frac{m^2_V}{m^2_V-m^2_W}\ln(\frac{m^2_V}{m^2_W}) \right)^2,
	\label{nuL->nuCL+J}
\end{eqnarray}
where $a=e,\mu,\tau$. $U_{ij}$ is the $U_{PMNS}$ neutrino mixing. For sake of simplicity we consider $U_{PMNS}$ real. Due to the cubic dependence of this width with the neutrino mass, we are going to have a large lifetime for this decay mode. In order to check how large it is, let us assume $v_\eta=v_\rho$  which implies, from the constraint $v^2_\eta+v^2_\rho=(247)^2$ GeV$^2$, that $v_\eta=v_\rho=176$GeV, and take   $m_{\nu_3}=0.2$eV. With those neutrino mixing angles presented in the beginning of this section, we obtain $\sum{U_{3 a}U_{a 1}}=U_{3 \tau}U_{\tau 1}+U_{3 \mu} U_{\mu 1}+U_{3e}U_{e 1}=0.7$. Using for the other parameters the values $g=0.66\,,\,\, m_W=81$GeV and taking $m_V=250$GeV\cite{boundM_V},
we obtain the following prediction to the  lifetime	of the neutrino decay mode $\nu_ {iL} \rightarrow (\nu^C)_{JL}+J$,
	\begin{eqnarray}
	\tau_{\nu_{3L} \rightarrow (\nu^C)_{1L}J}\approx 3\times10^{19}\mbox{s}.
	\label{nu3lifetime}
\end{eqnarray}
 Based on WMAP best fit, the age of the universe is $t_0=2.1\times10^{17}s$\cite{age}. This means that the decay mode $\nu_{3 L} \rightarrow (\nu^C)_{1 L}+J$ is  stable in face of the present age of the universe.

Let us then pass to the other  mode  where the daughter neutrino is a right-handed anti-neutrino, $\nu_{i L} \rightarrow (\nu^C)_{jR}+J$, as depicted in FIG.~(2). Here again we take $i=3$ and $j=1$. In order to see how such mode arises, note  that we can rewrite the charged current that couples with the charged gauge boson $W^+$ and $V^+$  in the following way,
\begin{eqnarray}
\frac{g}{\sqrt{2}}\bar e_L \gamma^\mu \nu_L W^+_\mu =\frac{g}{\sqrt{2}}\bar \nu^C_R \gamma^\mu e^C_R W^+_\mu .
	\label{rewritingCC}
\end{eqnarray}
In calculating that loop we find in the rest frame
\begin{eqnarray}
	\Gamma_{\nu_{3L} \rightarrow (\nu^C)_{1R}J}=\frac{ g^8(\displaystyle\sum_{a} U_{3a}m_aU_{a1})^2 v_\eta^2m_{\nu_3} }{ (16)^3\pi^5(m^2_{V}-m^2_{W})^2}\ln(\frac{m^2_V}{m^2_W})^2,
	\label{nuL->nuR+J}
\end{eqnarray} 
 Differently from the other mode, this one depends linearly on neutrino mass. Certainly this will imply a fast neutrino decay. To check this, let us substitute the same set of values for the parameters used in the  previous case, which leads to,
\begin{eqnarray}
	\tau_{\nu_{3L} \rightarrow (\nu^C)_{1R} J}\approx 8\times10^{-3}s\,.
	\label{nuR}
\end{eqnarray}
As expected we got a relatively fast neutrino decay mode. Observe that such prediction does not enter in conflict neither with the  SN1987A lower bound for this neutrino decay mode, which is, for $m_{\nu 3}=0.2$ eV, $\tau_{\nu3}\geq 2.5\times 10^{-8}$s~\cite{moha}, nor with the solar neutrino observations, whose lower bound for this case is $\tau_{\nu3}\geq 2\times 10^{-5}$s~\cite{beacom}

We know that fast neutrino decay has its main implications in cosmology and astrophysics. Such implications have already been  extensively investigated in the literature\cite{dolgov}. However all those investigations considered at least one heavy neutrino with mass around MeV's. After Super-kamiokande experiment have shown that the heavier active neutrino is around few eV´s,  the interest in neutrino decay in Majoron ceased because only a doublet or triplet Majoron (under 3-2-1) could lead to a fast decay. As we said before, such Majorons were ruled out~\cite{concha}. The main point of this section lies in the fact that our model is naturally providing a fast neutrino decay mode in Majoron that comes from a triplet by the 3-3-1 symmetry, and in this case the usual constraints that ruled out those Majorons do not apply here. This turns  this 3-3-1 model of particular interest from the point of view of neutrino cosmology.
%%%%%%%%%%%%%%%%%%%%%%%%%%%%%%%%%%%%%%%%%%%%%%%%%%%%%%%%%%%%%%%%%%%%%%%%%%%%%%%%%%%%%%
%                Figura decaimento de neutrino R.H, formato EPS.
%%%%%%%%%%%%%%%%%%%%%%%%%%%%%%%%%%%%%%%%%%%%%%%%%%%%%%%%%%%%%%%%%%%%%%%%%%%%%%%%%%%%%%
\begin{figure}[ht]
\begin{center}
\epsfig{file=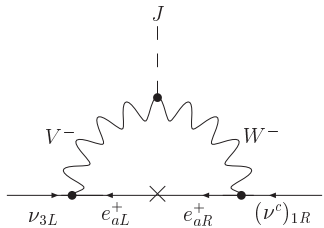,width=0.25\textwidth}
\caption{One loop diagram that leads to right-handed neutrino decay.}
\end{center}
\label{nuRfig}
\end{figure}
%%%%%%%%%%%%%%%%%%%%%%%%%%%%%%%%%%%%%%%%%%%%%%%%%%%%%%%%%%%%%%%%%%%%%%%%%%%%%%%%%%%%%%
%%%%%%%%%%%%%%%%%%%%%%%%%%%%%%%%%%%%%%%%%%%%%%%%%%%%%%%%%%%%%%%%%%%%%%%%%%%%%%
\section{Neutrinoless double beta decay. }
\label{sec3}
 Although the  experiments with neutrino oscillations showed that neutrinos are massive~\cite{SK,atm,solar,reactor,global}, they cannot determine if neutrinos are Majorana or Dirac particles. A crucial process that will help in determining neutrino nature is the neutrinoless double beta decay $(\beta\beta)_{0\nu}$\cite{experiment}. It is also a typical process which requires violation of the lepton number. Thus it can  be useful in  probing  new physics beyond the standard model.
 
Let us start our analysis reviewing the standard contribution. The interactions that lead to the  $(\beta\beta)_{0\nu}$ decay  involve  hadrons and leptons. For the case of the standard contribution, its amplitude  can be written as:
\begin{eqnarray}
	{\cal M}_{(\beta\beta)_{0\nu}}= \frac{g^4}{4m^4_W}{\cal M}^{(h)}_{\mu\nu}\bar u \gamma^\mu H_L \frac{q\!\!\!/ +m_\nu}{q^2-m_\nu^2}\gamma^\nu H_R v,
	\label{amplitud1}
\end{eqnarray}
with ${\cal M}^{(h)}_{\mu\nu}$ carrying the hadronic information of the process and $H_{R,L}=(1 \pm\gamma_5)/2$. In the presence of neutrino mixing and considering  that $m_\nu^2 << q^2$, thus we can write,
\begin{eqnarray}
	{\cal M}_{(\beta\beta)_{0\nu}}= {\cal A}_{(\beta\beta)_{0\nu}}{\cal M}^{(h)}_{\mu\nu}\bar u H_R\gamma^\mu  \gamma^\nu  v,
	\label{amplitud12}
\end{eqnarray}
with 
\begin{eqnarray}
{\cal A}_{(\beta\beta)_{0\nu}}=\frac{g^4 \langle M_\nu \rangle}{4m^4_W\langle q^2 \rangle}.
	\label{standardstrength}
\end{eqnarray}
being the  strength of the effective coupling of the standard contribution. For the case of three neutrino species   $\langle M_\nu \rangle = \sum U^2_{ei}m_{\nu_i}$ is the effective neutrino mass which we use as reference value $0.2$ eV. $\langle q^2 \rangle$ is the average of the transfered squared four-momentum. 

The  contributions to the  $(\beta\beta)_{0\nu}$ decay in our model come uniquely from the charged currents interactions with the charged gauge bosons $\tilde{W}^-$ and $\tilde{V}^-$. Perceive that, due to the mixing in Eq. (\ref{chargedmix}),  each  charged gauge boson  will interact, separately,  with the two leptonic charged currents $j^\mu$  and $j^{C \mu}$, as showed in Eq. (\ref{newCC}). Those new interactions generates a total of twelve contributions to the $(\beta\beta)_{0\nu}$ decay. Four  contributions involving only the  $\tilde{W}^-$, other four involving only the $\tilde{V}^-$ and other four involving both $\tilde{W}^-$ and $\tilde{V}^-$. 

In general, all the contributions to the $(\beta\beta)_{0\nu}$ decay will have an amplitude like this
\begin{eqnarray}
	{\cal M}_{(\beta\beta)_{0\nu}}= \frac{g^4}{4m^2_{X} m^2_{Y}}{\cal M}^{(h)}_{\mu\nu}\bar u\gamma^\mu H_L \frac{q\!\!\!/ +m_\nu}{q^2-m_\nu^2}\gamma^\nu H_R v,
	\label{amplitud2}
\end{eqnarray}
with $X^-$ and $Y^-$ being any of the charged gauge bosons $\tilde{W}^-$  and $\tilde{V}^-$. Note that Eq. (\ref{amplitud2}) implies that all the outgoing electrons are left-handed. After all this we can say that, except by the mixing angle in Eq. (\ref{chargedmix}), the effective couplings of any contribution to the $(\beta\beta)_{0\nu}$ process take then the form 
\begin{eqnarray}
{\cal A}_{(\beta\beta)_{0\nu}}(\mbox{new})=\frac{g^4 \langle M_\nu \rangle}{4 m^2_{X} m^2_{Y} \langle q^2 \rangle}.
	\label{newstrength}
\end{eqnarray}

As the $(\beta\beta)_{0\nu}$ decay has not been experimentally detected yet, the analysis we do here is to identify the new contributions and compare them with the standard one \cite{valle, mpp} in order to see if the model disposes of any  other relevant contribution  to the $(\beta\beta)_{0\nu}$ decay\cite{valle, mpp}.

To compare the contributions means to compare the effective couplings. Thus we first obtain the effective coupling of the standard contribution in our model, which involves only $\tilde{W}^-$ and the current $J^\mu$, as  depicted in FIG. (3). Its effective coupling takes now the form  
\begin{eqnarray}
{\cal A}_{(\beta\beta)_{0\nu}}(1)=\frac{g^4 \langle M_\nu \rangle}{4m^4_{\tilde{W}}\langle q^2 \rangle}c^4_\theta,
	\label{newstandardstrength}
\end{eqnarray}
%%%%%%%%%%%%%%%%%%%%%%%%%%%%%%%%%%%%%%%%%%%%%%%%%%%%%%%%%%%%%%%%%%%%%%
%        Figura 1 do decaimento beta  em EPS
%%%%%%%%%%%%%%%%%%%%%%%%%%%%%%%%%%%%%%%%%%%%%%%%%%%%%%%%%%%%%%%%%%%%%%
\begin{figure}[th]
\begin{center}
\epsfig{file=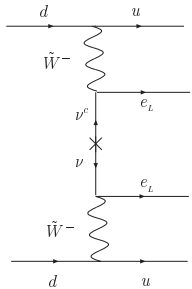,width=0.15\textwidth}
\caption{Standard contribuition to the $(\beta\beta)_{0\nu}$ decay.}
\end{center}
\end{figure}
%%%%%%%%%%%%%%%%%%%%%%%%%%%%%%%%%%%%%%%%%%%%%%%%%%%%%%%%%%%%%%%%%%%%%%%
%%%%%%%%%%%%%%%%%%%%%%%%%%%%%%%%%%%%%%%%%%%%%%%%%%%%%%%%%%%%%%%%%%%%%%%

The  first new contribution that we compare with the standard one is the contribution that involves  only  $\tilde{W}^-$ but now interacting with  the two charged currents, $J^\mu$  and $J^{C \mu}$.  Such a combination of currents generates two identical contributions that are depicted in  FIG. (4), whose effective coupling  is given by,
\begin{eqnarray}
{\cal A}_{(\beta\beta)_{0\nu}}(2)=\frac{g^4 \langle M_\nu \rangle}{4m^4_{\tilde{W}}\langle q^2 \rangle}c^3_\theta s_\theta.
	\label{strength1}
\end{eqnarray}
In comparing both effective couplings, we obtain the ratio
\begin{eqnarray}
	\frac{{\cal A}_{(\beta\beta)_{0\nu}}(2)}{{\cal A}_{(\beta\beta)_{0\nu}}(1)}=\tan\theta.
\label{comparison1}	
\end{eqnarray}
Note that the relevance of this contribution depends solely on the angle $\theta$. For $v_\eta=176$GeV  and $v_\chi=10^3$GeV, the $\rho$-parameter requires $v_{\eta^{\prime}}\leq 40$GeV\cite{majoron}. On taking  $v_{\eta^{\prime}}=40$GeV, we obtain 
\begin{eqnarray}
	\tan \theta\approx 0.007.
	\label{thata prediction}
\end{eqnarray}
This implies that 
\begin{eqnarray}
	{\cal A}_{(\beta\beta)_{0\nu}}(2)}=7\times 10^{-3}{\cal A}_{(\beta\beta)_{0\nu}(1).
\label{comparison1}	
\end{eqnarray}
This is  an interesting new contribution since that it mimics the  standard contribution once the particles involved are the same that appears in the standard contribution. Moreover the order of magnitude of this contribution is superior to those that involves new particles intermediating the process, as analyzed in Ref. \cite{valle}.

The other  contribution involving only $\tilde{W}^-$ is the one where the current $J^{C \mu}$ appears in the two vertices. In this case the ratio is proportional to $\tan^2\theta$. 
%%%%%%%%%%%%%%%%%%%%%%%%%%%%%%%%%%%%%%%%%%%%%%%%%%%%%%%%%%%%%%%%%%%%%%%%%%%%%%%%%%%%%%%%%%%%%%%%%%%
%                 Figura 2 do decaimento beta  em EPS
%%%%%%%%%%%%%%%%%%%%%%%%%%%%%%%%%%%%%%%%%%%%%%%%%%%%%%%%%%%%%%%%%%%%%%%%%%%%%%%%%%%%%%%%%%%%%%%%%%%
\begin{figure}[ht]
\begin{center}
\epsfig{file=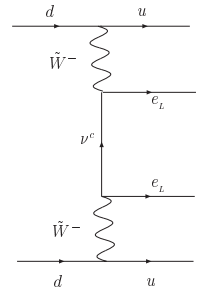,width=0.15\textwidth}
\caption{The same as the standard contribution but now involving the currents $J^\mu$  and $J^{C \mu}$.}
\end{center}
\end{figure}
%%%%%%%%%%%%%%%%%%%%%%%%%%%%%%%%%%%%%%%%%%%%%%%%%%%%%%%%%%%%%%%%%%%%%%%

Let us now consider  contributions that  involve both  $\tilde{W}^-$ and $\tilde{V}^-$. The first one we consider  is that  depicted in FIG.~(5). It involves the two currents $J^\mu$  and $J^{C \mu}$ interacting with $\tilde{W}^-$ and $\tilde{V}^-$, respectively. The  effective coupling in this case is given by,
\begin{eqnarray}
{\cal A}_{(\beta\beta)_{0\nu}}(3)=\frac{g^4 \langle M_\nu \rangle}{4m^2_{\tilde{W}}m^2_{\tilde{V}}\langle q^2 \rangle}c^3_\theta s_\theta.
	\label{strength2}
\end{eqnarray}	
Comparing with the standard effective coupling, we get the ratio
\begin{eqnarray}
\frac{{\cal A}_{(\beta\beta)_{0\nu}}(3)}{{\cal A}_{(\beta\beta)_{0\nu}}(1)}=\left( \frac{m_{\tilde{W}}}{m_{\tilde{V}}} \right)^2 \tan \theta.
\label{comparison2}	
\end{eqnarray}
Differently from the previous case, the relevance of this contribution depend on the angle $\theta$  and  the mass of $\tilde{V^-}$.  For $m_{\tilde{V}}=250$GeV, we obtain
\begin{eqnarray}
\frac{{\cal A}_{(\beta\beta)_{0\nu}}(3)}{{\cal A}_{(\beta\beta)_{0\nu}}(1)}\leq 7\times 10^{-4}. 
\label{comparison2prediction}	
\end{eqnarray}
For $m_{\tilde{V}}=\left( 500\,\,\mbox{and}\,\,1000\right)$ GeV  we get $\frac{{\cal A}_{(\beta\beta)_{0\nu}}(3)}{{\cal A}_{(\beta\beta)_{0\nu}}(1)}\leq 1.7\times10^{-4}$ and $4.7\times10^{-5}$, respectively. 

The next contributions are inferior in magnitude to those ones presented above, thus in what follow we just present their behavior with $\tan \theta$  and $m_{\tilde{V}}$.

The second possible contribution involving  $\tilde{V}^-$  and $\tilde{W}^-$ is the one where the current $ J^\mu$ appears in both vertices. In this case the ratio is proportional to $\left( \frac{m_{\tilde{W}}}{m_{\tilde{V}}} \right)^2 \tan^2 \theta$ which is equal to the contribution where  $J^\mu$ appears in both vertices instead of the current $J^{C \mu}$. The forth possible contribution in this case is the one where  the currents $J^\mu$  and $J^{C \mu}$ appear interacting with $\tilde{V}^-$ and $\tilde{W}^-$, respectively. The ratio now is proportional to $\left( \frac{m_{\tilde{W}}}{m_{\tilde{V}}} \right)^2 \tan^3 \theta$.

The last possible four contributions are the ones involving only $\tilde{V}^-$. In FIG. 6 we display an example of this kind of contribution where the current $J^{C \mu}$ appears in the two vertices. Its effective coupling is given by
\begin{eqnarray}
{\cal A}_{(\beta\beta)_{0\nu}}(4)=\frac{g^4 \langle M_\nu \rangle}{4m^4_{\tilde{V}}\langle q^2 \rangle}c^2_\theta s^2_\theta.
	\label{strength3}
\end{eqnarray}	
Comparing with the standard effective coupling, we get the ratio
\begin{eqnarray}
\frac{{\cal A}_{(\beta\beta)_{0\nu}}(4)}{{\cal A}_{(\beta\beta)_{0\nu}}(1)}=\left( \frac{m_{\tilde{W}}}{m_{\tilde{V}}} \right)^4 \tan^2 \theta.
\label{comparison3}	
\end{eqnarray} 

There are  two other contributions involving the currents $J^\mu$  and $J^{C \mu}$ and $\tilde{V}^-$. Both develop the same effective couplings and the ratio with the standard contribution get proportional to $\left( \frac{m_{\tilde{W}}}{m_{\tilde{V}}} \right)^4 \tan^3 \theta$. The fourth contribution with only $\tilde{V}^-$ is the one that involves only the current $J^\mu$. In this case the ratio with the standard contribution get proportional to $\left( \frac{m_{\tilde{W}}}{m_{\tilde{V}}} \right)^4 \tan^4 \theta$. 
%%%%%%%%%%%%%%%%%%%%%%%%%%%%%%%%%%%%%%%%%%%%%%%%%%%%%%%%%%%%%%%%%%%%%%%%%%%%%%%%%%%%%%%%%%%%%%%%%%
%                    Figura 3 decaimento beta 
%%%%%%%%%%%%%%%%%%%%%%%%%%%%%%%%%%%%%%%%%%%%%%%%%%%%%%%%%%%%%%%%%%%%%%%%%%%%%%%%%%%%%%%%%%%%%%%%%%
\begin{figure}[ht]
\begin{center}
\epsfig{file=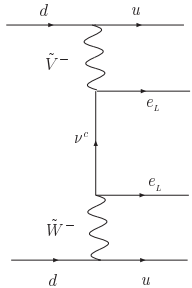,width=0.15\textwidth}
\caption{The main contribution involving both $\tilde{W}^-$ and  $\tilde{V}^-$.}
\end{center}
\end{figure}
%%%%%%%%%%%%%%%%%%%%%%%%%%%%%%%%%%%%%%%%%%%%%%%%%%%%%%%%%%%%%%%%%%%%%%%%%%%%%%%%%%%%%%%%%

Let us briefly discuss neutrinoless double beta decay with the emission of one Majoron. We refer to this decay as $(\beta \beta)_{0\nu J}$. Despite the fact that our Majoron does not couple directly with neutrinos, the $(\beta \beta)_{0\nu J}$ decay is still possible due to the coupling of the Majoron with the charged gauge bosons  given in Eq.~(\ref{W+V+J}). In face of the mixing in Eq.~(\ref{chargedmix}), the coupling in Eq.~(\ref{W+V+J}) gets multiplied by $c^2_\theta$.  

The dominant contribution for the $(\beta \beta)_{0\nu J}$ decay  is the one depicted in FIG.~(7), with an effective coupling given by 
\begin{eqnarray}
{\cal A}_{(\beta\beta)_{0\nu J}}=\frac{ g^6 v_\eta \langle M_\nu \rangle c_\theta^6} {8m^4_{\tilde{W}}m^2_{\tilde{V}}\langle q^2 \rangle }.
	\label{strengthJ}
\end{eqnarray}	
Comparing with the standard effective coupling, we get the ratio
\begin{eqnarray}
\frac{{\cal A}_{(\beta\beta)_{0\nu J}}}{{\cal A}_{(\beta\beta)_{0\nu}}(1)}=\frac{ g^2 v_\eta   c^2_\theta Q} {2m^2_{\tilde{V}}},
\label{comparison4}	
\end{eqnarray}
where $Q$ denotes the energy available in the process and whose value is $2038.56$ KeV~\cite{Q}. In our analysis here we take $Q=1$MeV. For $g=0.66$, $m_{\tilde{V}}=250$ GeV, $v_\eta=247/\sqrt{2}$ GeV and taking $c_\theta=1$, we get,
\begin{eqnarray}
\frac{ {\cal A}_{(\beta\beta)_{0\nu J}}}{{\cal A}_{(\beta\beta)_{0\nu}}(1)}\leq6.0\times 10^{-5}.
\label{comparison3prediction}	
\end{eqnarray} 
This contribution is in the level of magnitude of that one involving one $\tilde{V}^+$ depicted in FIG. (5). 
%%%%%%%%%%%%%%%%%%%%%%%%%%%%%%%%%%%%%%%%%%%%%%%%%%%%%%%%%%%%%%%%%%%%%%%%%%%%%%%%%%%%%%%%%%%%%%%%%%
%                    Figura 3 decaimento beta 
%%%%%%%%%%%%%%%%%%%%%%%%%%%%%%%%%%%%%%%%%%%%%%%%%%%%%%%%%%%%%%%%%%%%%%%%%%%%%%%%%%%%%%%%%%%%%%%%%%
\begin{figure}[ht]
\begin{center}
\epsfig{file=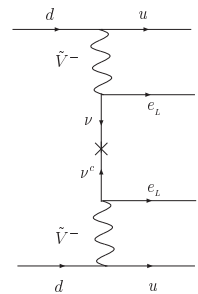,width=0.15\textwidth}
\caption{The main contribution involving two $\tilde{V}^-$.}
\end{center}
\end{figure}
%%%%%%%%%%%%%%%%%%%%%%%%%%%%%%%%%%%%%%%%%%%%%%%%%%%%%%%%%%%%%%%%%%%%%%%%%%%%%%%%%%%%%%%%%%%%%%%%%% 
\section{conclusions}
\label{sec4}
In this paper we investigated the implications of spontaneous breaking of the lepton number in the 3-3-1 model with right-handed neutrinos in  neutrino decay with Majoron emission and in neutrinoless double beta decay with and without Majoron emission. 

In regard to neutrino decay with Majoron emission, the main result is that, although this decay arises only through radiative correction, the model still disposes of a fast neutrino decay mode, namely $\nu_{3L} \rightarrow (\nu^C)_{1R} +J$, with lifetime of order of $8\times10^{-3}$s, which is of the order of magnitude to be interesting for cosmological purposes.

With respect to neutrinoless double beta decay, the spontaneous breaking of the lepton number generates a proliferation of contributions.  Differently from other cases\cite{mpp,valle,mohabeta}, all the  new contributions take place with the exchange of virtual light neutrinos, similarly to the standard contribution. We performed a systematic analysis of the couplings  of all contributions. The relevance of the new contributions will be dictated by the   mixing angle $\theta$ and the mass $m_{\tilde V}$. Thus, to determine correctly the value of the angle $\theta$  and the mass $m_{\tilde V}$ turns out to be a matter of primary interest in what concerns the relevance of the new contributions to the neutrinoless double beta decay in the model in question. By estimating the order of  magnitude of the  new contributions, we predicted that the most robust one is that depicted in FIG. (4) whose order of magnitude is $7\times 10^{-3}$ of the standard contribution. If these new contributions are important or not,  only the accuracy of the experiments will tell us.

In summary,  the spontaneous breaking of the lepton number in the 3-3-1 model with right-handed neutrinos leads to fast neutrino decay in Majoron and generates a bunch of new  contributions to the neutrinoless double beta  decay whose relevance is dictated by the angle $\theta$ and the mass $m_{\tilde V}$.

%%%%%%%%%%%%%%%%%%%%%%%%%%%%%%%%%%%%%%%%%%%%%%%%%%%%%%%%%%%%%%%%%%%%%%%%%%%%%%%%%%%%%%%%%%%%%%%%%%
\begin{figure}[ht]
\begin{center}
\epsfig{file=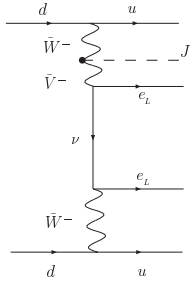,width=0.17\textwidth}
\caption{The main contribution to the $(\beta\beta)_{0\nu J}$ decay.}
\end{center}
\end{figure}

%%%%%%%%%%%%%%%%%%%%%%%%%%%%%%%%%%%%%%%%%%%%%%%%%%%%%%%%%%%%%%%%%%%%%%%%%%%%%%%

{\it Acknowledgments.}  This work was supported by Funda\c c\~ao de
Amparo \`a Pesquisa do Estado de S\~ao Paulo (FAPESP)(AD,AGD),
Conselho Nacional de Desenvolvimento Cient\'{\i}fico e
Tecnol\'ogico (CNPq) (CASP,PSRS) and  by Funda\c c\~ao de
Apoio \`a Pesquisa do Estado da Para\'{\i}ba(FAPESQ) (CASP).
%%%%%%%%%%%%%%%%%%%%%%%%%%%%%%%%%%%%%%%%%%%%%%%%%%%%%%%%%%%%%%%%%%%%%%%%%%%%%%%


\begin{thebibliography}{99}
\bibitem{coment}
 We have to stress that, although we dispose in Eq. (\ref{leptonmass}) of more free parameters than constraints, it is not straightforward to generate the correct spectrum of lepton masses and mixing. This is so because of the complexity of $m^\nu_{ij}$. A numerical approach becomes necessary. In a work in progress we are implementing a FORTRAN program in order to obtain all the possible spectrum of neutrino masses and mixings generated by the simultaneous diagonalization of those matrices.
\end{thebibliography}
\end{document}